# Local algorithms for the prime factorization of strong product graphs

Marc Hellmuth, Wilfried Imrich, Werner Klöckl and Peter F. Stadler

**Abstract.** The practical application of graph prime factorization algorithms is limited in practice by unavoidable noise in the data. A first step towards error-tolerant "approximate" prime factorization, is the development of local approaches that cover the graph by factorizable patches and then use this information to derive global factors. We present here a local, quasi-linear algorithm for the prime factorization of "locally unrefined" graphs with respect to the strong product. To this end we introduce the *backbone* $\mathbb{B}(G)$ for a given graph $G$ and show that the neighborhoods of the backbone vertices provide enough information to determine the global prime factors.

## 1. Introduction

Network structures derived from real-life data are notoriously incomplete and/or plagued by measurement errors. As a consequence, the structures need to be analyzed in a way that is robust against inaccuracies, noise, and perturbations in the data. Most of the classical graph invariants, however, are very fragile in this respect. The property of factorizability w.r.t. to each of the associative graphs products [5] is no exception. In fact, very small perturbations already can destroy the product structure completely, modifying a graph with many factors to a prime one [2, 9]. The recognition of product structures, even if disguised by noise, is a problem of practical interest e.g. in theoretical biology [8] and in mathematical engineering [7]. A first systematic investigation into "approximate graph products" showed, that a practically viable approach can be based on *local* factorization algorithms, because these recognize product structures in small subgraphs and attempt to stepwisely extend regions with product structures. This idea has been fruitful in particular for the strong product of graphs, where one benefits from the fact that the local product structure of neighborhoods is a refinement of the global factors [4].

In [4] we have introduced the class of *thin-neighborhood intersection coverable (NICE)* graphs, and devised an efficient quasi-linear-time local factorization algorithm w.r.t. the strong product. Here we extend this previous work to the much larger class of thin graphs which are whose local factorization is not finer than the global one. Will call this property *locally unrefined.*

Fig. 4 gives an example of a thin but locally refined graph. It is prime but the closed neighborhoods of all vertices have two prime factors. We can regard this graph as an approximate product, arising from a product of paths that are then connected in a twisted way that destroys factorizability at the global level but leaves the local product structure intact. A local approach like the one proposed here, will recognize the regional product structure and only at a later stage report global inconsistencies. A detailed discussion of this point is, however, beyond the scope of this contribution. Here, we shall focus only on the recognition of globally factorizable graphs by means of a novel, strictly local approach. It is based on the concept of the *backbone* $\mathbb{B}(G)$ of a thin graph, defined on the cardinality of equivalence classes of a particular relation $S$. As it turns out $\mathbb{B}(G)$ is the set of vertices with strictly maximal closed neighborhoods. We will show how to determine the prime factors of a given locally unrefined graph $G$ by covering it by neighborhoods of the backbone vertices only. Moreover, we will derive polynomial-time local algorithms for computing the product coloring and the Cartesian skeleton of $G$, and for recognizing whether $G$ is locally unrefined.

## 2. Background and Results

### 2.1. Definitions

**Basic Notation.** We only consider finite, simple, connected and undirected graphs $G = (V, E)$ with vertex set $V$ and edge set $E$. A graph is *nontrivial* if it has at least two vertices. The neighborhood $N(v)$ of a vertex $v \in V$ is the set of all vertices that are adjacent to $v$. Throughout this contribution, we will mostly be concerned with closed neighborhoods $N[v] := N(v) \cup \{v\}$. We define the *n-neighborhood* of vertex $v$ as the set $N_n[v] = \{x \in V(G) \mid d(v, x) \leq n\}$, where $d(x, y)$ denotes the canonical distance in $G$, i.e., the length of a shortest path connecting the vertices $x$ and $y$. Notice that $N_1[v] = N[v]$. Unless there is a risk of confusion, we call a 1-neighborhood just neighborhood. The degree $deg(v)$ of a vertex $v$ is the number adjacent vertices, or, equivalently, the number of incident edges. The maximum degree in a given graph is denoted by $\Delta$. The subgraph of a graph $G$ that is induced by a vertex set $W \subseteq V(G)$ is denoted by $\langle W \rangle$.

**Product Graphs.** The vertex set of the *strong product* $G_1 \boxtimes G_2$ of two graphs $G_1$ and $G_2$ is defined as $\{(v_1, v_2) \mid v_1 \in V(G_1), v_2 \in V(G_2)\}$, the Cartesian product of the vertex sets of the factors. Two vertices $(x_1, x_2), (y_1, y_2)$ are adjacent in $G_1 \boxtimes G_2$ if one of the following conditions is satisfied:

(i) $[x_1, y_1] \in E(G_1)$ and $[x_2, y_2] \in E(G_2)$,

(ii) $[x_1, y_1] \in E(G_1)$ and $x_2 = y_2$,

(iii) $[x_2, y_2] \in E(G_2)$ and $x_1 = y_1$.

The *Cartesian product* $G_1 \square G_2$ has the same vertex set as $G_1 \boxtimes G_2$, but vertices are only adjacent if they satisfy (ii) or (iii). Consequently, edges that satisfy (ii) or (iii) in a strong product are called Cartesian, the other edges non-Cartesian. In products of more factors, the edges that join vertices differing in exactly one coordinate are called Cartesian.

It is well known that the strong product is associative. Thus a vertex of $\boxtimes G_i$ is properly "coordinatized" by the vector $(x_1, \ldots, x_n)$ whose entries are the vertices $x_i$ of its factor graphs $G_i$. The coordinatization of a product is equivalent to a (partial) edge coloring of $G$ in which edges $(xy)$ share the same color $c_k$ if the two points $x$ and $y$ differ only in the value of a single coordinate $k$, i.e., if $x_i = y_i$, $i \neq k$ and $x_k \neq y_k$. This colors the *Cartesian edges* of $G$ (with respect to the *given* product representation). It follows that for each color $c$ the set $E_c = \{e \in E \mid c(e) = c\}$ of edges with color $c$ spans $G$. The connected components of $\langle E_c \rangle$, usually called the *layers* or *fibers* of $G$, are isomorphic subgraphs of $G$.

For later reference we note that the distance of two vertices and product graph is determined by distances within the factors:

**Lemma 2.1 ([5], p.149).** *Let $G = G_1 \boxtimes \cdots \boxtimes G_k$ be the strong product of connected graphs. Then*

$$d_G(u, v) = \max_{1 \leq i \leq k} d_{G_i}(u_i, v_i).$$

A *subproduct* of a product $G \boxtimes H$ is defined as the product of subgraphs of $G$ and $H$ respectively. In particular, neighborhoods are subproducts:

**Lemma 2.2 ([4]).** *For any two graphs $G$ and $H$ holds $\langle N_n^{G \boxtimes H}[(x, y)] \rangle = \langle N_n^G[x] \rangle \boxtimes \langle N_n^H[y] \rangle$.*

Let $G = \boxtimes_{j=1}^n G_j$ be a strong product graph and denote the coordinates of $x \in V(G)$ by $(x_1, \ldots x_{i-1}, x_i, x_{i+1}, \ldots, x_n)$. The mapping $p_i(x) = x_i$ is called *projection* of $x$ onto the $i - th$ factor. $G_i^x$ denotes the subgraph that is induced by the vertices of the set $\{(x_1, \ldots x_{i-1}, v, x_{i+1}, \ldots, x_n) \in V(G) \mid v \in V(G_i)\}$. We call this subgraph $G_i^x$ a $G_i$-*fiber* or $G_i$-*layer* through vertex $x$. With a *horizontal* fiber we mean the subgraph of $G$ induced by vertices of one and the same fiber, i.e. we mean a particular $G_i^x$-fiber without mention this particularly, if there is no risk of confusion. With *parallel* $G_i$-fibers we mean all fibers with respect to a given factor $G_i$. Edges of (not necessarily) different $G_i$-fibers are said to be edges *of one and the same* factor $G_i$. If a subgraph $H$ of $G = G_1 \boxtimes G_2$ has a representation $H = H_1 \square H_2$ that is compatible with the strong product $G$ in the sense that $H_i$-fibers and the $G_i$-fibers induce the same partition of the vertex sets $V(H) = V(G)$, we call the subgraph $H$ *Cartesian skeleton*, see [5].

A graph $G$ is *prime* with respect to the strong product if it is not isomorphic to the strong product of two nontrivial graphs. It is clear that every finite graph is isomorphic to a strong product of prime graphs. This product representation usually is called prime factor decomposition (PFD).

For finite connected graphs, the prime factorization with respect to the strong product is unique up to isomorphism and the order of the factors [1]. There are, however, disconnected graphs with a non-unique PFD, see e.g. [5].

**Thinness.** An important issue in the context of graph products is whether or not two vertices can be distinguished by their neighborhoods. This is captured by the relation $S$, see [1]: Two vertices $x$, $y \in V(G)$ are in the relation $S$ if they have the same closed neighborhoods, $N[x] = N[y]$. Note that $x S y$ implies that $x$ and $y$ are adjacent since, by definition, $x \in N[x]$ and $y \in N[y]$. Clearly, $S$ is an equivalence relation. The $S$-class of $x$ is denoted by $S(x) = \{v \in V \mid N[v] = N[x]\}$.

The graph $G/S$ is the usual quotient graph. More precisely, its vertex set is $V(G/S) = \{S_i \mid S_i \text{ is an equivalence class of } S\}$ and $(S_i, S_j) \in E(G/S)$ whenever $(x, y) \in E(G)$ for some $x \in S_i$ and $y \in S_j$. Note that $S_{G/S}$ is trivial, that is, its equivalence classes are single vertices [5]. We call a graph $S$-*thin*, or *thin* for short, if it has trivial $S$, i.e. if no two vertices are in relation $S$. Thus $G/S$ is thin. The importance of thinness lies in the following uniqueness result:

**Theorem 2.3 ([1, 5]).** *A finite, undirected, simple, connected graph $G$ has a unique PFD with respect to the strong product. If $G$ is $S$-thin, then the coordinatization is unique.*

As a consequence, the Cartesian edges are uniquely determined in $S$-thin graphs. Note that the set of all Cartesian edges in a thin strong product $G$ are the edges of the Cartesian skeleton $H$ of $G$. An example of a non $S$-thin graph and with different possible coordinatizations is shown in Figure 1.

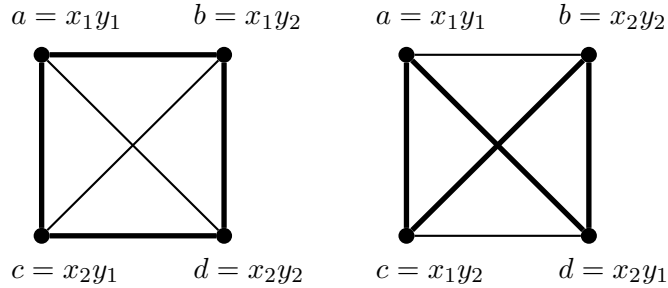

FIGURE 1. The edge (a,b) is Cartesian in the left and non-Cartesian in the right coordinatization.

Much of our work is based on the intuition that it should be feasible to construct the Cartesian skeleton of $G$ by considering only PFDs of suitable neighborhoods, more precisely, it should be sufficient to work with the Cartesian edges of these neighborhoods. The main problem with this idea is that it requires that $G$ can be covered by thin neighborhoods $\langle N[v] \rangle$, a condition that is not always met.

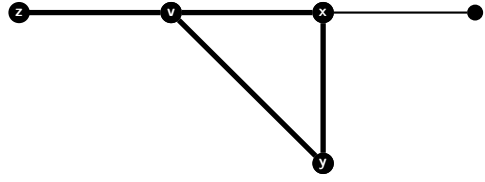

FIGURE 2. A thin graph whereby $\langle N[v] \rangle$ is not thin. The S-classes in $\langle N[v] \rangle$ are $S_v(v) = \{v\}$, $S_v(z) = \{z\}$ and $S_v(x) = S_v(y) = \{x, y\}$.

The main obstacle is the realization that even though $G$ is thin, this is not necessarily true for subgraphs, and in particular neighborhoods, Fig. 2. In order to investigate this issue in some more detail, we also consider $S$-classes w.r.t. subgraphs of a given graph $G$.

**Definition 2.4.** Let $H \subseteq G$ be an arbitrary subgraph of a given graph $G$. Then $S_H(x)$ is defined as the set

$$S_H(x) = \{v \in V(H) \mid N[v] \cap V(H) = N[x] \cap V(H)\}$$

For the sake of convenience we set $S_y(x) := S_{\langle N[y] \rangle}(x) = \{v \in N[y] \mid N[v] \cap N[y] = N[x] \cap N[y]\}$. In other words, $S_y(x)$ is the $S$-class that contains $x$ in the subgraph $\langle N[y] \rangle$. Notice that $N[x] \subseteq N[v]$ holds for all $v \in S_x(x)$. If $G$ is additionally thin, then $N[x] \subsetneq N[v]$.

**Lemma 2.5 ([5], p.155).** *For any two graphs $G_1$ and $G_2$ holds $(G_1 \boxtimes G_2)/S \simeq G_1/S \boxtimes G_2/S$. Furthermore, for every $x = (x_1, x_2) \in V(G)$, $S_G(x) = S_{G_1}(x_1) \times S_{G_2}(x_2)$.*

**Corollary 2.6.** *Let $G$ be a strong product $G = G_1 \boxtimes G_2$. Consider a vertex $x \in V(G)$ with coordinates $(x_1, x_2)$. Then for every $z \in S_G(x)$ holds $z_i \in S_{G_i}(x_i)$, i.e. the $i$-th coordinate of $z$ is contained in the $S$-class of the $i$-th coordinate of $x$.*

*Moreover, $G$ and $G/S$ have the same number of nontrivial prime factors if and only if none of the prime factors of $G$ is isomorphic to a complete graph.*

**The S1-condition.** Since the Cartesian edges are globally uniquely defined in a thin graph, the challenge is to find a way to determine enough Cartesian edges from local information, even if $\langle N[v] \rangle$ is not thin. The following property will play a crucial role for this purpose:

**Definition 2.7.** Let $G = (V, E)$ be a thin graph. An edge $(x, y) \in E$ satisfies the *S1-condition* if there is vertex $z \in V$ s.t.

1. $x, y \in N[z]$ and

2. $|S_z(x)| = 1$ or $|S_z(y)| = 1$.

From Lemma 2.5 we can directly infer that the cardinality of an $S$-class in a product graph $G$ is the product of the cardinalities of the corresponding $S$-classes in the factors. Applying this fact together with Lemma 2.2 to the subgraph of $G$ induced by a closed neighborhood $N[v]$ immediately implies Corollary 2.8.

**Corollary 2.8.** *Consider a strong product $G = G_1 \boxtimes G_2$ and two vertices $v, x \in V(G)$ with coordinates $(v_1, v_2)$ and $(x_1, x_2)$, s.t. $v_i, x_i \in V(G_i)$ and $v_i \in N[x_i]$ for $i = 1, 2$. Then $S_v(x) = S_{v_1}(x_1) \times S_{v_2}(x_2)$ and therefore $|S_v(x)| = |S_{v_1}(x_1)| \cdot |S_{v_2}(x_2)|$.*

**Lemma 2.9.** *Let $G$ be a strong product graph containing two $S$-classes $S_G(x)$, $S_G(y)$ that satisfy*

*(i) $(S_G(x), S_G(y))$ is a Cartesian edge in $G/S$ and*
*(ii) $|S_G(x)| = 1$ or $|S_G(y)| = 1$.*

*Then all edges in $G$ induced by vertices of $S_G(x)$ and $S_G(y)$ are Cartesian and copies of one and the same factor.*

*Proof.* For simplicity, we write $S(\,.\,)$ for $S_G(\,.\,)$. Suppose the PFD of $G$ consists of $n$ factors. Furthermore, we may assume w.l.o.g. that $|S(x)| = 1$. Lemma 2.5 implies that for every prime factor $G_i$ of $G$, $1 \leq i \leq n$, holds

$$|S_{G_i}(x_i)| = 1$$

In the following, $S(v)_m$ denotes the $m$-th coordinate of vertex $S(v)$ in $G/S$. Being a Cartesian edge means that $S(x)$ and $S(y)$ coincide in every, but one, say the $j$-th coordinate w.r.t. the PFD of $G/S$, i.e. $\forall i \neq j$ holds $S(x)_i = S(y)_i$. By Lemma 2.5 this is $S_{G_i}(x_i) = S_{G_i}(y_i)$.

Corollary 2.6 implies that the $i$-th coordinate ($i \neq j$) of every vertex in $S(x) \cup S(y)$ is in $S_{G_i}(x_i) \cup S_{G_i}(y_i) = S_{G_i}(x_i)$, which is a set of cardinality 1. Hence, all vertices in $S(x) \cup S(y)$ have the same $i$-th coordinate. This is equivalent to the claim of the lemma. □

*Remark* 2.10. Whenever we find a Cartesian edge $(x, y)$ in a neighborhood $\langle N[z] \rangle$ so that one endpoint of $(x, y)$ is contained in a $S$-class of cardinality 1 in $\langle N[z] \rangle / S$, i.e., such that $S_z(x) = \{x\}$ or $S_z(y) = \{y\}$, we can therefore conclude that all edges in $\langle N[z] \rangle$ induced by vertices of $S_z(x)$ and $S_z(y)$ are also Cartesian and are copies of one and the same factor.

**The Backbone of $G$.** This observation naturally gives rise to two questions:

- Which Cartesian edges have these properties?
- What is the global structure of a Cartesian skeleton that can be determined in this way?

We first consider a subset of $V(G)$ that will play a special role in our approach.

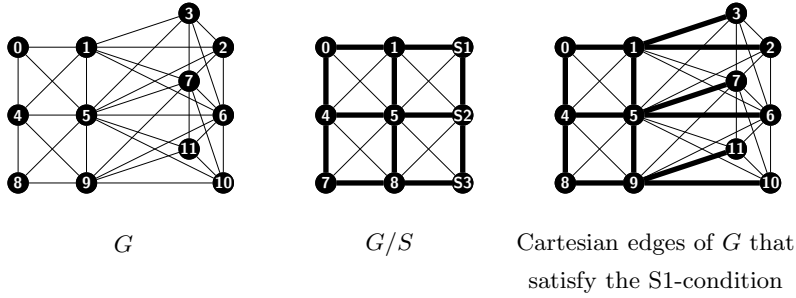

$G$          $G/S$          Cartesian edges of $G$ that
satisfy the S1-condition

FIGURE 3. This figure shows how we identify (a couple of) Cartesian edges that satisfy the *S1-condition*. Given a graph $G$ compute its quotient graph $G/S$, decompose $G/S$ w.r.t. to its strong prime factors. Since $G/S$ is thin the Cartesian edges in $G/S$ are uniquely determined. Apply now Lemma 2.9 to identify all Cartesian edges in $G$ that satisfy the *S1-condition* in $G/S$.

**Definition 2.11.** The *backbone* of a thin graph $G$ is the vertex set
$$\mathbb{B}(G) = \{v \in V(G) \mid |S_v(v)| = 1\}.$$

Notice that Corollary 2.6 implies that the number of prime factors of $\langle N[v] \rangle$ coincides with the number of prime factors of $\langle N[v]/S \rangle$ for all $v \in \mathbb{B}(G)$, otherwise we would have $|S_v(v)| > 1$.

**Definition 2.12.** Let $G = (V, E)$ be a graph. A subset $D$ of $V$ is a *dominating set* for G, if for all vertices in $V \setminus D$ there is at least one adjacent vertex from $D$.
We call $D$ *connected dominating set*, if $D$ is a dominating set and the subgraph induced by $D$ is connected.

## 2.2. Main Results

In this subsection we give preview of our main results. The proofs will then be given in the next sections.

**Theorem 2.13.** *Let $G$ be a thin graph. Then the backbone $\mathbb{B}(G)$ is a connected dominating set for $G$.*

For all $v \in V(G)$, there exists a vertex $w \in N[v]$ s.t. $w \in \mathbb{B}(G)$. In other words, the entire graph can be covered by closed neighborhoods of the backbone vertices. The following result highlights the importance of $\mathbb{B}(G)$ for the identification of Cartesian edges.

**Theorem 2.14.** *All Cartesian edges that satisfy the* S1-condition *in an arbitrary induced neighborhood also satisfy the* S1-condition *in the induced neighborhood of*

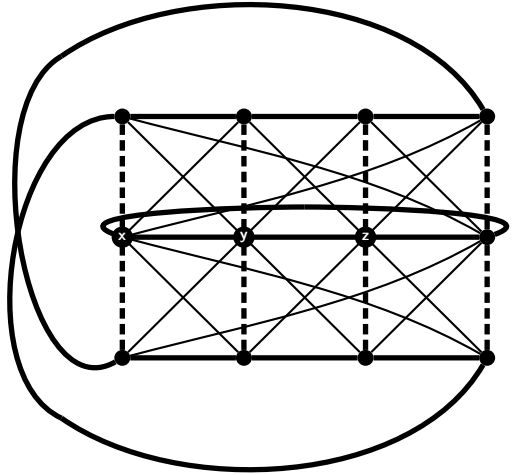

FIGURE 4. A prime graph where each local factorization leads to
two factors. This graph is often also called a *twisted* product.

*a vertex of the backbone* $\mathbb{B}(G)$.
*If at least one edge of* $G_i^x$ *in* $G = \boxtimes_{j=1}^n G_j$ *satisfies the* S1-condition *then all
vertices of* $G_i^x$ *are contained in edges of* $G_i^x$ *that satisfy the* S1-condition.

This result implies that it makes sense to give the following definition:

**Definition 2.15.** An entire $G_i^x$-fiber satisfies the *S1-condition*, whenever one of its
edges does.

Taken together, these two theorems allow us to identify all Cartesian edges
of $G_i^x$-fiber that satisfies the *S1-condition*, using exclusively information about the
neighborhoods of the backbone vertices.

The identification of the Cartesian edges is only the first step in constructing
a PFD, however. We also need to be able to determine which Cartesian edges
belong to copies of the same factor $G_i$. This task is explained in the next sections.
As we shall see, the PFD can be *computed* with particular efficiency for graphs in
which the local PDF is not finer than the global one.

**Definition 2.16.** Let $G$ be a given graph and $H \subseteq G$ be an arbitrary subgraph.
$|PF(G)|$ and $|PF(H)|$ denote the number of the prime factors of $G$ and $H$, respectively.
The graph class $\Upsilon$ of *locally unrefined graphs* consists of all $S$-thin graphs with the
property that $|PF(G)| = |PF(\langle N[v] \rangle)|$ for all $v \in \mathbb{B}(G)$.
The graph class $\Upsilon_n$ is the set of all graphs $G \in \Upsilon$ with $|PF(G)| = n$.

Not all graphs have this property. A counter-example is shown in Figure 4.

The following theorems state the time complexity of our developed algorithms working on a local level and concerning the PFD and the determination of the Cartesian skeleton of graph $G \in \Upsilon$ and the recognition if $G \in \Upsilon$.

**Theorem 2.17 (PFD of $G \in \Upsilon$).** *Let $G = (V, E) = \boxtimes_{j=1}^n G_j \in \Upsilon$ with bounded maximum degree $\Delta$. Then the prime factors of $G$ can be determined in $O(|V| \cdot \log_2(\Delta) \cdot \Delta^6)$ time.*

**Theorem 2.18.** *Let $G = (V, E) \in \Upsilon$ with bounded maximum degree $\Delta$. Then the Cartesian skeleton of $G$ can be determined and colored w.r.t. to the prime factors in $O(|V|^2 \cdot \Delta^{10})$ time.*

**Theorem 2.19.** *Let $G = (V, E)$ be a thin graph with bounded maximum degree $\Delta$. Determining whether $G \in \Upsilon$ can be done in $O(|V|^2 \cdot \Delta^{10})$ time.*

## 3. Backbone

We start exploring properties of the backbone $\mathbb{B}(G)$ of thin graphs. Our immediate goal is to establish that every vertex of $G$ is in $\mathbb{B}(G)$ or has an adjacent vertex contained in $\mathbb{B}(G)$. This is equivalent to the statement that $\mathbb{B}(G)$ is a domination set for $G$. We will then show that $\mathbb{B}(G)$ is *connected*.

**Lemma 3.1.** *Let $G$ be a thin, connected simple graph and $v \in V(G)$ with $|S_v(v)| > 1$. Then there exists a vertex $y \in S_v(v)$ s.t. $|S_y(y)| = 1$.*

*Proof.* Let $|S_v(v)| > 1$. Since $G$ is finite we can choose a vertex $y \in S_v(v)$ that has a maximal closed neighborhood in $G$ among all vertices in $S_v(v)$. Moreover $N[y]$ is maximal in $G$ among all vertices of $V(G)$. Assume not. Then there is a vertex $z$ s.t. $N[y] \subset N[z]$, but then $z \in S_v(v)$, a contradiction to the maximality of $N[y]$ among all vertices in $S_v(v)$. Since $G$ is thin $N[y]$ is strictly maximal.

Furthermore $|S_y(y)| = 1$, otherwise there is a $z \in S_y(y)$, $z \neq y$ s.t. $N[z] \cap N[y] = N[y]$. Since $G$ is thin, there is a $x \in N[z]$ with $x \notin N[y]$ and thus $N[y] \subsetneq N[z]$, but this is a contradiction to the fact that $N[y]$ is strictly maximal. □

**Lemma 3.2.** *Let $G$ be a thin graph and $v$ an arbitrary vertex of $G$. Then $v \in \mathbb{B}(G)$ if and only if $N[v]$ is a strictly maximal neighborhood in $G$*

*Proof.* If $N[v]$ is a strictly maximal neighborhood in $G$ then $|S_v(v)| = 1$ which is shown analogously to the last part of the last proof.

Let now $v \in \mathbb{B}(G)$. Assume $N[v]$ is not strictly maximal. Then there is a vertex $z \in V(G)$ different from $v$ such that $N[v] \subseteq N[z]$. Thus, $N[v] \cap N[z] = N[v]$, $z \in S_v(v)$ and $|S_v(v)| > 1$, contradicting that $v \in \mathbb{B}(G)$. □

**Lemma 3.3.** *Let $G$ be a thin connected simple graph. Then the backbone $\mathbb{B}(G)$ is a dominating set for $G$.*

*Proof.* We have to show that for all $v \in V(G)$ there exists a vertex $w \in N[v]$ s.t. $|S_w(w)| = 1$. If $\langle N[v] \rangle$ is thin or $|S_v(v)| = 1$, there is nothing to show. If $|S_v(v)| > 1$, then the statement follows from Lemma 3.1. □

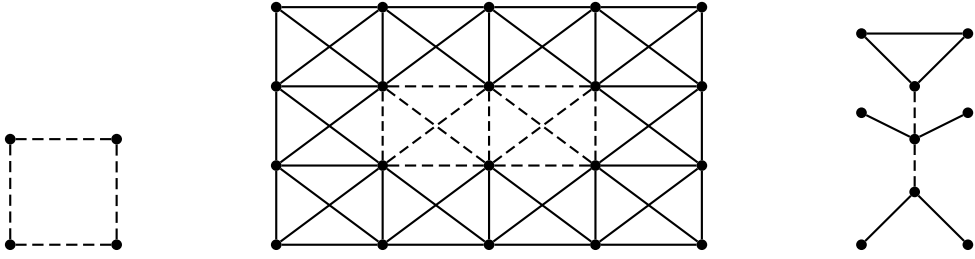

Figure 5. Examples of backbones, highlighted by the dashed lines.

**Lemma 3.4.** *Let $G$ be a thin connected simple graph. Then the set of adjacent vertices $v$ and $w$ with $|S_w(w)| = 1$ or $|S_v(v)| = 1$ induces* one *connected subgraph $H$ of $G$.*

*Proof.* Assume $H$ consists of at least two components and let $\mathcal{C}$ denote the set of these components. Since $G$ is connected we can choose components $C, C' \in \mathcal{C}$ s.t. there are vertices $x \in C$, $y \in C'$ that are adjacent in $G$. Since $G$ is finite and $x, y \in N[x]$ there is a maximal closed neighborhood $N[z]$ in $G$ containing $x$ and $y$. The thinness of $G$ implies that $N[z]$ is strictly maximal. This implies, analogously as in the proof of Lemma 3.1, that $|S_z(z)| = 1$ contradicting that $x$ and $y$ are in different components of $H$. □

**Lemma 3.5.** *Let $G$ be a thin connected graph. Then the set of adjacent vertices $v$ and $w$ with $|S_w(w)| = 1$ and $|S_v(v)| = 1$ induces* one *connected subgraph $H$ of $G$, i.e. the backbone $\mathbb{B}(G)$ induces a connected subgraph $H$ of $G$.*

*Proof.* Assume $H$ consists of at least two connected component. Let $C$ be any such connected component. From Lemma 3.4 we can conclude that the subgraph $M$ of $G$ induced by all vertices of edges $(v, w)$ with $|S_w(w)| = 1$ or $|S_v(v)| = 1$ is connected. Hence, in $M$ there is path $P = \{x = x_0, x_1, x_2, ..., x_{n-1}, x_n = y\}$ from $x \in C$ to $y \in C'$, where $C'$ is any other connected component.

W.l.o.g., we may assume that $P \cap V(C) = \{x\}$. (Otherwise we replace $P$ by $\{x_m, x_{m+1}, ..., x_n = y\}$, where $m = \max\{i \mid x_i \in P \cap V(C)\}$.) This implies that $x_1$ is not in $\mathbb{B}(G)$. But then $x_2$ must be in a component $C'' \neq C$ from $\mathbb{B}(G)$, since every edge in $M$ contains at least one vertex which is in $\mathbb{B}(G)$.

Notice that neither $x$ nor $x_2$ are in $S_{x_1}(x_1)$, otherwise $(x, x_2) \in E(G)$ and $C$ and $C''$ would be connected. By Lemma 3.1 we can choose a $z \in S_{x_1}(x_1), z \neq x, x_2$ with $|S_z(z)| = 1$. Thus $C$ and $C''$ are connected. Contradiction. □

From Lemma 3.3 and Lemma 3.5 we can directly infer Theorem 2.13. Notice that if $G$ is thin and $|\mathbb{B}(G)| > 1$, then *every* vertex has an adjacent vertex that is in the backbone. Clearly this is not true whenever $|\mathbb{B}(G)| = 1$, as the example in Figure 6 shows.

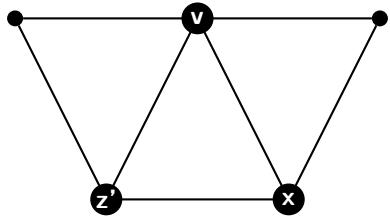

Figure 6. A thin graph $G$ with backbone $\mathbb{B}(G) = \{v\}$. Thus there is no vertex $w \in N(v)$ s.t. $|S_w(w)| = 1$. Moreover notice that $|S_{z'}(x)| = 1$ but $x, z' \notin \mathbb{B}(G)$. Lemma 3.7 implies that there is a vertex $z \in \mathbb{B}(G)$ such that $|S_z(x)| = 1$. In this example holds $z = v$.

**Lemma 3.6.** *Let $G$ be a thin graph with a backbone consisting of a single vertex $\mathbb{B}(G) = \{v\}$. Then $|S_v(w)| = 1$ for all $w \in V(G)$.*

*Proof.* Theorem 2.13 implies that $\langle N[v] \rangle \simeq G$ and thus $S_v(w) = S_G(w)$ for all $w \in V(G)$. Since $G$ is thin every $S$ class in $G$ is trivial and therefore also in $\langle N[v] \rangle$. $\qquad\square$

The backbone vertices play a special role in our approach. First we observe that all Cartesian edges $(x, y)$ that have a vertex which is contained in the backbone satisfy the *S1-condition*. Furthermore, we will show next that it suffices to consider only induced neighborhoods of backbone vertices to find all Cartesian edges that satisfy the *S1-condition*.

**Lemma 3.7.** *Let $G$ be a thin graph and $(x, y)$ an arbitrary edge in $E(G)$. If there exists a vertex $z' \in N[x] \cap N[y]$ with $|S_{z'}(x)| = 1$ then there exists even a vertex $z \in N[x] \cap N[y]$ with the following properties:*

$$z \in \mathbb{B}(G) \ and \ |S_z(x)| = 1.$$

*Proof.* If $z' \in \mathbb{B}(G)$ there is nothing to show.
Now suppose $|S_{z'}(z')| > 1$. By Lemma 3.1 we can choose a vertex $z \in S_{z'}(z')$ with $|S_z(z)| = 1$. Since $z \in S_{z'}(z')$, we can conclude that $N[z'] \subset N[z]$ and thus $x, y \in N[z]$ and therefore $z \in N[x] \cap N[y]$.
It remains to show that $|S_z(x)| = 1$. Assume $|S_z(x)| > 1$ then there is a vertex $w \in S_z(x)$ different from $x$. The definition of $S_z(x)$ implies $N[w] \cap N[z] = N[x] \cap N[z]$, which implies that $w \in N[z']$, since $z' \in N[x] \cap N[z]$. Moreover we can conclude

$$N[w] \cap N[z] \cap N[z'] = N[x] \cap N[z] \cap N[z']. \tag{3.1}$$

Since $N[z'] \subset N[z]$, we can cancel the intersection with $N[z]$ in equation 3.1 to obtain

$$N[w] \cap N[z'] = N[x] \cap N[z'].$$

But then $w \in S_{z'}(x)$ and thus $|S_{z'}(x)| > 1$, contradicting $|S_{z'}(x)| = 1$. Hence $|S_z(x)| = 1$. □

**Lemma 3.8.** *Let* $(x, y) \in E(G)$ *be an arbitrary edge in a thin graph* $G$ *such that* $|S_x(x)| > 1$*. Then there exists a vertex* $z \in \mathbb{B}(G)$ *s.t.* $z \in N[x] \cap N[y]$*.*

*Proof.* Since $|S_x(x)| > 1$ and by applying Lemma 3.1 we can choose a vertex $z \in S_x(x)$ with $z \in \mathbb{B}(G)$. Since $z \in S_x(x)$ it holds $N[x] \subset N[z]$ and hence $y \in N[z]$, and the claim follows. □

**Corollary 3.9.** *Let* $G$ *be a thin graph and* $(x, y)$ *an arbitrary edge in* $E(G)$ *that does not satisfy the* S1-*condition. Then there exists a vertex* $z \in \mathbb{B}(G)$ *s.t.* $z \in N[x] \cap N[y]$*, i.e. the edges* $(z, x)$ *and* $(z, y)$ *satisfy the* S1-*condition.*

## 4. Determining the prime factors of $G \in \Upsilon$ in quasi-linear time

In this section we will first show that it is possible to determine all (Cartesian) edges of a $G_i^x$-fiber, whenever one edge of this $G_i^x$-fiber satisfies the *S1-condition*. Then we will explain how we can color such fibers in a way that all edges of a particular $G_i^x$-fiber receive the same color, restricted to the fact that $x \in \mathbb{B}(G)$. The general case will be treated in Section 5. Moreover we will give an algorithm that determines the prime factors of a given graph $G \in \Upsilon$ in quasi-linear time in the number of vertices of $G$.

### 4.1. Determining Cartesian edges

First we will prove the following. If at least one edge of a fiber $G_i^x$ satisfies the *S1-condition* in $G$, then all vertices contained in $G_i^x$ have an endpoint in an edge $e \in E(G_i^x)$ that satisfies the *S1-condition*. In practice, this means that we can mark *all* edges of the subgraph $G_i^x$ as Cartesian if at least one edge of $G_i^x$ satisfies the *S1-condition* provided that the corresponding product graph $G$ is thin.

**Lemma 4.1.** *Let* $G = \boxtimes_{j=1}^n G_j$ *be the strong product of thin graphs and* $(x, y) \in E(G)$ *be a Cartesian edge, where* $x$ *and* $y$ *differ in coordinate* $i$*. Moreover let* $(x, y)$ *satisfy the* S1-*condition. Then for all edges* $(a, b) \in E(G_i^x)$ *at least one of the following statements is true:*

1. $(a, b)$ *satisfies the* S1-*condition.*
2. *There are edges* $(\tilde{z}, a), (\tilde{z}, b) \in E(G_i^x)$ *that satisfy the* S1-*condition.*
   *In this case, knowing that* $(\tilde{z}, a), (\tilde{z}, b)$ *belong to* $G_i^x$ *implies that* $(a, b)$ *is necessarily also an edge of* $G_i^x$*.*

*Furthermore, the vertices incident with edges of* $G_i^x$ *that satisfy the* S1-*condition induce a single connected subgraph* $H \subseteq G_i^x$*.*

*Proof.* By associativity and commutativity of the strong product it suffices to show this for the product $G = G_1 \boxtimes G_2$ of two thin (not necessarily prime) graphs. Notice that $G_i^x = G_i^y$, since $x$ and $y$ differ only in coordinate $i$. Furthermore let $(x_1, x_2)$ denote the coordinates of $x$. The notation of the coordinates of $a$, $b$, and $y$ is analogous. W.l.o.g. assume $i = 2$ and $|S_z(x)| = 1$ with $z = (z_1, z_2) \in N[x] \cap N[y]$. Corollary 2.8 implies $|S_{z_1}(x_1)| = 1$ and $|S_{z_2}(x_2)| = 1$. The idea of the rest of the proof is to shift properties of $(a_2, b_2)$, the projection of $(a, b)$ into the factor $G_2$, to $(a, b)$.

Case (a) $(a_2, b_2)$ satisfies the *S1-condition* in $G_2$. Then we may assume w.l.o.g. that there is a $v_2 \in G_2$ with $|S_{v_2}(a_2)| = 1$ and $a_2, b_2 \in N[v_2]$. Since $x_1 = a_1$, Corollary 2.8 implies $|S_{(z_1, v_2)}(a)| = 1$. Lemma 2.1 shows that $a, b \in N[(z_1, v_2)]$. This means $(a, b)$ satisfies the *S1-condition*.

Case (b) $(a_2, b_2)$ does not satisfy the *S1-condition* in $G_2$. Then Corollary 3.9 implies the existence of a vertex $v_2 \in G_2$ such that both $(v_2, a_2)$ and $(v_2, b_2)$ satisfy the *S1-condition* in $G_2$. Case (a) shows that $((a_1, v_2), a)$ and $((a_1, v_2), b)$ satisfy the *S1-condition*.

Since $\mathbb{B}(G_2)$ is a connected dominating set for $G_2$, the subgraph of $G_2$ induced by all vertices of edges that satisfy the *S1-condition* in $G_2$ is connected. Since we can shift every edge that satisfies the *S1-condition* in $G_2$ to an edge that satisfies the *S1-condition* in $G_i^x$, $H$ is connected.                                   $\square$

From Lemma 3.7 and 4.1 we can directly conclude that Theorem 2.14 holds.

Clearly, we can identify at least one edge of each prime factor that belongs to the backbone of $G$, even if the local decomposition is finer than the global one. In the case of locally unrefined graphs $G \in \Upsilon$, however, we can do much more: Once we have found an edge $(x, y)$ of a $G_i^x$-fiber that satisfies the *S1-condition* we can identify *all* edges of that $G_i^x$-fiber as Cartesian.

It remains to show how we can identify edges not only as Cartesian but also as belonging to a fiber of a particular factor $G_i$. In the next subsection we will explain how this works

### 4.2. Identify colors of $G_i^x$-fibers with $x \in \mathbb{B}(G)$

In this subsection we will show how to color a $G_i^x$-fiber of a given product graph $G \in \Upsilon$ with $x \in \mathbb{B}(G)$ in a way that all edges of the $G_i^x$-fiber receive the same color. For this we will need several definitions that are quite similar to those ones in [4]. Here, we will restrict the constructions to individual $G_i^x$-fibers, however.

**Definition 4.2.** A *product coloring* of a graph $G = \boxtimes_{i=1}^n G_i$ is a mapping $F_G$ from a subset $E'$ of the set of Cartesian edges of $G$ into a set $C = \{1, \ldots, n\}$ of colors, such that all edges in a $G_i$-fibers receive the same color $i$.

Let $G_j^x$ be a fiber of an arbitrary factor $G_j$ of $G$. An *$(x, j)$-partial product coloring $((x, j)$-PPC)* of a graph $G = \boxtimes_{i=1}^n G_i$ is a mapping $F_G$ from a subset $E'$ of the set of Cartesian edges of $G$ into a set $C$ of colors, such that all edges in this particular $G_j^x$-fiber receive the same color.

**Definition 4.3.** Let $H_1, H_2 \subset G$ and $F_{H_1}$ be a $(x, j)$-PPC of $H_1$. Then $F_{H_2}$ is a $(x, j)$-*color continuation* of $F_{H_1}$ if there is a color $c$ in the image of $F_{H_2}$ such that there is an edge of the $G_j^x$-fiber that satisfies the *S1-condition* in $H_2$ with color $c$ that is also in the domain of $F_{H_1}$ and satisfies the *S1-condition* in $H_1$. More formally:

$$\exists \text{ edge } e \in Dom(F_{H_1}) \cap Dom(F_{H_2}) \cap E(G_j^x)$$

that satisfies the *S1-condition* in both $H_1$ and $H_2$.

The *combined* $(x, j)$-*PPC* on $H_1 \cup H_2$ uses the color of $F_{H_1}$ on $H_1$ and colors all edges $f$ of $H_2$ with $F_{H_2}(f) = c$ with the color $F_{H_1}(e)$.

**Definition 4.4.** A finite sequence $\sigma_{(x,j)} = (v_i)_{i=0}^k$ of vertices of $G$ is a $(x, j)$-*covering sequence* if

1. for all $v \in V(G_j^x)$ there exists a vertex $w \in \sigma_{(x,j)}$ with $v \in N[w]$ and
2. if for all $i > 0$ every PPC of $\langle N[v_{i+1}] \rangle$ is a $(x, j)$-color continuation of the combined $(x, j)$-coloring of $\bigcup_{l=1}^i E(\langle N[v_l] \rangle)$ defined by the $(x, j)$-PPC of each $\langle N[v_l] \rangle$.

We call a $(x, j)$-*covering sequence* simply *covering sequence* if there is no risk of confusion.

In order to construct a proper covering sequence we introduce an ordering of the vertices $V = \{w_0, \ldots, w_{n-1}\}$ of $G$ by means of breadth-first search: Select an arbitrary vertex $v \in V$ and create a sorted list $BFS(v)$ of vertices beginning with $v$; append all neighbors $v_1, \ldots, v_{d(v)}$ of $v$; then append all neighbors of $v_1$ that are not already in this list; continue recursively with $v_2, v_3, \ldots$ until all vertices of $V$ are processed. In this way we build levels where each $v$ in level $i$ is adjacent to some vertex $w$ in level $i - 1$ and vertices $u$ in level $i + 1$. We then call the vertex $w$ the *parent* of $v$, denoted by $parent(v)$, and vertex $v$ a *child* of $w$.

In our approach we will use this search algorithm in a slightly modified way. Let $v \in \mathbb{B}(G)$ be the starting vertex. We then decompose the neighborhood of $v$ w.r.t. to its strong prime factor decomposition. Then we fix one color $c$ of one fiber, say $G_i^v$, and append only those neighbors $v_j$ of $v$ to the current list $BFS(v)$ if

1. they are not already in this list and
2. $v_j \in \mathbb{B}(G)$ and
3. the edge $(v, v_j)$ has the color $c$ of the corresponding $G_i^v$-fiber.

This will be done recursively for the remaining vertices $w$ fixing the color in each neighborhood $\langle N[w] \rangle$ of the underlying $G_i^v$-fiber. Therefore, $BFS(v)$ is a sorted $BFS$-list on the vertex set $\mathbb{B}(G) \cap V(G_i^v)$.

First we show that in a prime graph $G \in \Upsilon$ such a $BFS(x)$ ordering on the vertices of $\mathbb{B}(G)$ leads to a $(x, 1)$-covering sequence of $G$.

**Lemma 4.5.** *Let $G \in \Upsilon$ be prime and let $x$ be an arbitrary vertex of the backbone $\mathbb{B}(G) = \{w_1, \ldots, w_m\}$. Then $BFS(x)$ on the vertices of $\mathbb{B}(G)$ is a $(x, 1)$-covering sequence.*

*Proof.* By Theorem 2.13 holds that for all $v \in V(G)$ there is a vertex $w \in BFS(x)$ such that $v \in N[w]$. Thus item (1) of Definition 4.4 is fulfilled.

Notice that $|PF(\langle N[v] \rangle)| = 1$ for all $v \in BFS(x)$ since $G \in \Upsilon$. Thus all edges in such $\langle N[v] \rangle$ are Cartesian and get exactly one color.

Now, take two arbitrary consecutive vertices $v_i, v_{i+1}$ from $BFS(x)$. If $v_i$ and $v_{i+1}$ are adjacent then $v_{i+1}$ is a child of $v_i$ and the edge $(v_i, v_{i+1})$ satisfies the *S1-condition* in $\langle N[v_i] \rangle$ as well as in $\langle N[v_{i+1}] \rangle$, since $v_i, v_{i+1} \in \mathbb{B}(G)$. Therefore the edge $(v_i, v_{i+1})$ is colored in the neighborhoods of both adjacent vertices and we get a proper $(x, 1)$-color continuation from $\langle N[x] \rangle \cup \bigcup_{l=1}^{i} \langle N[v_l] \rangle$ to $\langle N[v_{i+1}] \rangle$.

If $v_i$ and $v_{i+1}$ are not adjacent (thus $v_i \neq x$) then there must be parents $u, w \in BFS(x)$ of $v_i$ and $v_{i+1}$, resp. and we can apply the latter argument. Therefore $BFS(x)$ is a proper $(x, 1)$-covering sequence. □

We will now directly transfer that knowledge to (non prime) product graphs. For this we will introduce in Algorithm 1 how to get a proper coloring on all $G_i^x$-fiber with $x \in \mathbb{B}(G)$. The correctness is proved in the following lemma. Remind that $\Upsilon_n \subset \Upsilon$ denotes the set of graphs $G \in \Upsilon$ with $|PF(G)| = n$.

**Lemma 4.6.** *Let $G \in \Upsilon_n$ and $x$ be an arbitrary vertex of $\mathbb{B}(G)$. Then Algorithm 1 properly colors all edges of each $G_i^x$-fiber for $i = 1, \ldots, n$.*

*Proof.* We show in the sequel that the $BFS$ covering of vertices of $\mathbb{B}(G) \cap V(G_i^x)$, i.e of vertices along Cartesian edges $(a, b)$ of $G_i^x$ with $a, b \in \mathbb{B}(G)$, leads to a proper $(x, i)$-covering sequence.

First notice that for each $x \in \mathbb{B}(G)$ holds $|PF(\langle N[x] \rangle)| = |PF(G)| = n$, since $G \in \Upsilon_n$. Moreover, all Cartesian edges $(v, w)$ with $v, w \in \mathbb{B}(G) \cap V(G_i^x)$ satisfy the *S1-condition* and therefore can be determined as Cartesian, by applying Lemma 2.2 and Lemma 2.9. Hence, any such edge $(v, w)$ was properly colored both in $\langle N[w] \rangle$ and in $\langle N[v] \rangle$. Applying Theorem 2.3 leads to the requested PPC.

We show next that for all vertices $y \in G_i^x$ there is a vertex $w \in N[y]$ with $w \in BFS(x)$, implying that item (1) of Definition 4.4 is fulfilled. Since $\mathbb{B}(G_i)$ is a connected dominating set for factor $G_i$ we can conclude that for all vertices $y_i \in V(G_i)$ there is a vertex $w_i \in N[y_i]$ such that $w_i \in \mathbb{B}(G_i)$. Suppose that the coordinates for the chosen vertices $x$ are $(x_1, \ldots, x_n)$. Then, $w \in N[y]$ has coordinates $(x_1, \ldots, x_{i-1}, w_i, x_{i+1}, \ldots, x_n)$. Corollary 2.8 implies $|S_{x_j}(x_j)| = 1$ for $j = 1, \ldots, n$ Furthermore, we have $S_w(w) = \prod_{j=1}^{i-1} |S_{x_j}(x_j)| \cdot |S_{w_i}(w_i)| \cdot \prod_{j=i+1}^{n} |S_{x_j}(x_j)| = 1$. Thus $w \in V(G_i^x) \cap \mathbb{B}(G)$ and therefore $w \in BFS(x)$.

Moreover, since all those edges $(w, y)$ with $y \notin \mathbb{B}(G)$ satisfy the *S1-condition* and the fact that $G \in \Upsilon_n$ we can conclude that these edges are properly colored in the neighborhood $\langle N[w] \rangle$. Therefore $BFS(x)$ along vertices of $\mathbb{B}(G) \cap V(G_i^x)$ constitutes a proper $(x, i)$-covering sequence $\sigma_{x,i}$.

Finally, consider Line $17 - 22$ of the algorithm. Theorem 2.13 and Lemma 4.1 imply that the remaining edges $(y, y')$ of $G_i^x$ that do not satisfy the *S1-condition* are induced by vertices of Cartesian edges $(z, y)$ and $(z, y')$ that do satisfy the *S1-condition*. As shown above, all those edges $(z, y)$ and $(z, y')$ are already colored

---

**Algorithm 1** Color $G_i^x$-fiber

---

1: **INPUT:** a thin graph $G \in \Upsilon_n$ and a vertex $x \in \mathbb{B}(G)$
2: factor $\langle N[x]/S \rangle$ and properly color the Cartesian edges in $\langle N[x] \rangle$ that satisfy the *S1-condition* with colors $c_1, \ldots, c_n$;
3: $L_i \leftarrow \emptyset, i = 1, \ldots, n$;
4: **for** $i = 1, \ldots, n$ **do**
5:     mark $x$;
6:     add all neighbors $v \in \mathbb{B}(G)$ of $x$ with color $c_i$ in list $L_i$ in the order of their covering;
7:     **while** $L_i \neq \emptyset$ **do**
8:         take first vertex $v$ from the front of $L_i$;
9:         delete $v$ from $L_i$;
10:        **if** $v$ is not marked **then**
11:            mark $v$;
12:            factor $\langle N[v]/S \rangle$ and properly color the Cartesian edges in $\langle N[v] \rangle$ that satisfy the *S1-condition*;
13:            combine the colors on edge $(parent(v), v)$;
14:            add all neighbors $w \in \mathbb{B}(G)$ of $v$ with color $c_i$ to the end of list $L_i$ in the order of their covering;
15:        **end if**
16:    **end while**
17:    **for** all edges $(v, w)$ that do not satisfy the *S1-condition* **do**
18:        **if** there are edges $(z, v)$ and $(z, w)$ that have color $c_i$ **then**
19:            mark $(v, w)$ as Cartesian and assign color $c_i$ to $(v, w)$;
20:            {Notice that these edges $(z, v)$ and $(z, w)$ satisfy the *S1-condition*}
21:        **end if**
22:    **end for**
23: **end for**
24: **OUTPUT:** G with colored $G_j^x$-fiber, $j = 1, \ldots, n$;
25: {Notice that every $G_j^x$-fiber is isomorphic to one prime factor of $G$}

---

with the same color in some $\langle N[w] \rangle$ with $w \in V(G_i^x) \cap \mathbb{B}(G)$. It follows that we obtain a complete coloring in $G_i^x$.

   This procedure is repeated independently for all colors $c_i$ in $\langle N[x] \rangle$, $i = 1, \ldots, n$. This completes the proof.          $\square$

**Lemma 4.7.** *Algorithm* 1 *determines the prime factors of a given graph $G = (V, E) \in \Upsilon$ with bounded maximum degree $\Delta$ in time complexity $O(|V| \cdot \log_2(\Delta) \cdot (\Delta)^6)$.*

*Proof.* The time complexity of Algorithm 1 is determined by the complexity of the BFS search and the decomposition of each neighborhood in each step.
Notice that the number of vertices of every neighborhood $N[v]$ is at most $\Delta + 1$.

With the algorithm of Feigenbaum and Schäffer [3] it can be factored in $O((\Delta + 1)^5)$ time. The number of colors is bounded by the number of factors in each neighborhood, which is at most $\log_2(\Delta+1)$. The BFS search takes at most $O(|V| + |E|)$ time for each color. Since the number of edges in $G$ is bounded by $|V| \cdot \Delta$ we can conclude the the time complexity of the BFS search is $O(|V| + |V| \cdot \Delta) = O(|V| \cdot \Delta)$. Thus we end in an overall time complexity of $O((|V| \cdot \Delta) \cdot \log_2(\Delta + 1) \cdot (\Delta + 1)^5)$ which is $O(|V| \cdot \log_2(\Delta) \cdot (\Delta)^6)$. □

If $G \in \Upsilon$, it is sufficient to use Algorithm 1 to identify a single $G_i$-fiber through exactly one vertex $x \in \mathbb{B}(G)$ in order to determine the corresponding prime factor of $G$. For $G \in \Upsilon$ we would therefore be ready at this point. Finding a vertex that is in the backbone can be efficiently done and thus its cost can be ignored in the estimation of the complexity in Theorem 2.17 (see the proof of Lemma 5.8). Hence together with Lemma 4.6 and Lemma 4.7 we can directly conclude Theorem 2.17.

There is, however, no known sufficient condition to establish that $G \in \Upsilon$, except of course by computing the PFD of $G$. Moreover, as discussed in [4], it will be very helpful to determine as many identificable fibers as possible for applications to approximate graph products. However, this task will be treated in Section 5.

## 5. Detection and product coloring of the Cartesian skeleton

As shown in the last section we can identify and even color $G_i^x$-fiber that satisfy the *S1-condition* in a way that all edges of this fiber receive the same color, whenever $x \in \mathbb{B}(G)$. We will generalize this result for all fibers that satisfy the *S1-condition* in Lemma 5.1. This provides that we get a big part of the Cartesian skeleton colored such that all edges of identified $G_i^y$-fibers received the same color. Moreover we will show how to identify colors of different colored $G_i$-fibers. Furthermore we introduce a "brute-force" method to determine Cartesian edges of fibers that do not satisfy the *S1-condition*.

### 5.1. Identify colors of all $G_i^x$-fibers that satisfy the S1-condition

**Lemma 5.1.** *Let $G \in \Upsilon_n$ and $G_i^y$ with $y \notin \mathbb{B}(G)$ be an arbitrary fiber that satisfies the* S1-condition. *Let $z \in \mathbb{B}(G)$ such that $|S_z(a)| = 1$ or $|S_z(b)| = 1$ for some edge $(a, b) \in G_i^y$. Then the $(z, i)$-covering sequence $\sigma_{z,i}$ is also a $(y, i)$-covering sequence.*

*Proof.* The existence of such a vertex $z$ follows directly from Lemma 3.7. W.l.o.g. let $|S_z(a)| = 1$, otherwise switch the labels of vertices $a$ and $b$. If $G_i^y = G_i^z$ then the assumption follows directly from Lemma 4.6. Thus we can assume that $G_i^y \neq G_i^z$.

W.l.o.g. let vertex $z$ have coordinates $(z_1, \ldots, z_i, \ldots, z_n)$ and vertex $a$ have coordinates $(a_1, \ldots, a_i, \ldots, a_n)$. In the following $\hat{z}$ will denote the vertex in $G_i^y$ with coordinates $\hat{z}_j = a_j$ for $j \neq i$ and $\hat{z}_i = z_i$, in short with coordinates $(a_1, \ldots, z_i, \ldots, a_n)$. Thus we can infer that $G_i^{\hat{z}} = G_i^y$. For the sake of convenience we will denote all vertices with coordinates $(z_1, \ldots, w_i, \ldots, z_n)$ and $(a_1, \ldots, w_i, \ldots, a_n)$ with $w$ and $\hat{w}$, respectively. Remark that $w$ and $\hat{w}$ are adjacent, by choice of their coordinates and by definition of the strong product.

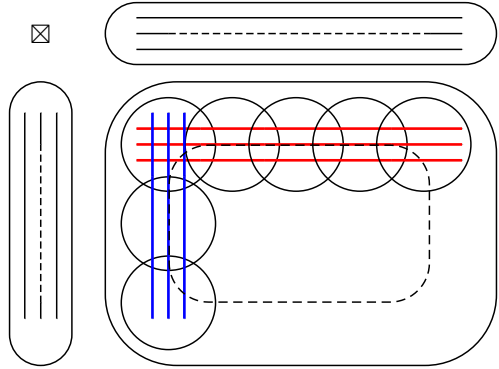

FIGURE 7. The Backbone is depicted as dashed line. Starting with some vertex $x \in \mathbb{B}(G)$ we go along backbone vertices of $G$ with fixed color, i.e. we apply the BFS algorithm only on vertices $\mathbb{B}(G) \cap G_i^x$ for all $i$. Applying Lemma 4.6, 5.1 and 5.2 we can color all $G_i$-fibers that satisfy the *S1-condition*.

Moreover, since $a \in N[z]$ and because of the coordinates of the vertices $\hat{u}, \hat{w} \in G_i^y$ we can infer that $\hat{u} \in N[w]$ holds for all vertices $\hat{u} \in N[\hat{w}]$ by definition of the strong product. More formally, $N[\hat{w}] \cap V(G_i^y) \subseteq N[w]$, see Figure 8.

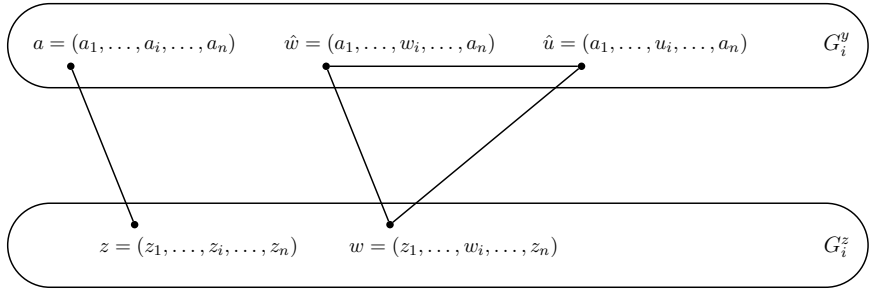

FIGURE 8. $N[\hat{w}] \cap V(G_i^y) \subseteq N[w]$

Let $\sigma_{z,i} = (z, v^1, \ldots, v^m)$ be a proper $(z, i)$-covering sequence, based on the BFS approach explained above, consisting of all backbone vertices of $G$ contained in $G_i^z$. Furthermore let $w$ be any vertex of $\sigma_{z,i}$. Notice that for all such vertices $w$ holds $|S_w(w)| = 1$ and therefore in particular $|S_{w_i}(w_i)| = 1$, by applying Corollary

2.8. Thus for all such vertices $\hat{w}$ holds

$$|S_w(\hat{w})| = \prod_{j=1}^{i-1} |S_{z_i}(a_i)| \cdot |S_{w_i}(w_i)| \cdot \prod_{j=i+1}^{n} |S_{z_i}(a_i)| = 1,$$

by applying Corollary 2.8 again. Hence all edges $(\hat{u}, \hat{w}) \in E(\langle N[\hat{w}]\rangle) \cap E(G_i^y)$ satisfy the *S1-condition* in the closed induced neighborhood of the vertex $w$, since $N[\hat{w}] \cap V(G_i^y) \subseteq N[w]$. Moreover since $\mathbb{B}(G_i)$ is a connected dominating set we can infer that item (1) of Definition 4.4 is fulfilled.

It remains to show that we also get a proper color continuation. The main challenge now is to show that for all vertices $(parent(v), v)$ contained in $BFS(z)$ there is an edge $(a, b) \in G_i^y$ that satisfies the *S1-condition* in both $\langle N[parent(v)]\rangle$ and $\langle N[v]\rangle$. This implies that we can continue the color of the $G_i^y$-fiber on that edge $(a, b)$.

Therefore, let $\hat{v}$ and $\hat{w}$ be any two adjacent vertices of $G_i^y$ with coordinates as mentioned above such that $v_i, w_i \in \mathbb{B}(G_i)$. Thus by choice of the coordinates $v$ and $w$ are adjacent vertices such that $|S_v(v)| = |S_w(w)| = 1$ and hence $v, w \in BFS(z)$. As shown above $|S_v(\hat{v})| = 1$ and $|S_w(\hat{w})| = 1$. Therefore the edge $(\hat{v}, \hat{w})$ satisfies the *S1-condition* in both $\langle N[\hat{v}]\rangle$ and $\langle N[\hat{w}]\rangle$, since $N[\hat{v}] \cap V(G_i^y) \subseteq N[v]$ and $N[\hat{w}] \cap V(G_i^y) \subseteq N[w]$. The connectedness of $\mathbb{B}(G_i)$ and $G \in \Upsilon_n$ implies that any such edge is properly colored with $c$ by means of the color continuation. Since $\mathbb{B}(G_i)$ is also a dominating set it holds that all vertices $\hat{u}$ with $|S_u(\hat{u})| > 1$ have an adjacent vertex $\hat{w}$ with $|S_w(\hat{w})| = 1$. Since $N[\hat{w}] \cap V(G_i^y) \subseteq N[w]$, we can infer that $\hat{u} \in N[w]$ and therefore all these edges satisfy the *S1-condition* and are colored with $c$. Hence, property (2) of Definition 4.4 is satisfied. □

**Lemma 5.2.** *Let $G \in \Upsilon_n$ and $G_i^y$ with $y \notin \mathbb{B}(G)$ be an arbitrary fiber that satisfies the* S1-condition. *Furthermore let $z \in \mathbb{B}(G)$ with $|S_z(a)| = 1$ or $|S_z(b)| = 1$ for some edge $(a, b) \in G_i^y$. Then Algorithm 1 properly colors all edges of each such $G_i^y$-fiber with vertex $z$ as an input vertex.*

*Proof.* Lemma 3.7 implies that there is a $z \in \mathbb{B}(G)$ such that $|S_z(a)| = 1$ or $|S_z(b)| = 1$ for some edge $(a, b) \in G_i^y$. As shown in Lemma 5.1 each such $G_i^y$-fiber that satisfies the *S1-condition* with $y \notin \mathbb{B}(G)$ can be covered and colored via the corresponding $(z, i)$-covering sequence $\sigma_{z,i}$. By the way since $G \in \Upsilon_n$ and Theorem 2.3 we can directly color all edges of $G_i^y$ with the same color $c$ as the $G_i^z$-fiber. Furthermore applying Lemma 4.1 all remaining edges of $(a, b) \in E(G_i^y)$ are induced by vertices of Cartesian edges $(a, \tilde{z}), (b, \tilde{z}) \in E(G_i^y)$ which are satisfying the *S1-condition* and thus already colored with color $c$. Thus all these edges $(a, b)$ must be Cartesian edges of $G_i^y$ (by definition of the strong product) and thus also obtain color $c$. □

## 5.2. Identification of parallel fibers

As shown in the last subsection we are able to identify all edges of a $G_i^x$-fiber that satisfies the *S1-condition* as Cartesian in such a way that all these edges in $G_i^x$ get same color. An example of the colored Cartesian edges of a product graph after coloring all horizontal fibers that satisfy the *S1-condition* is shown in Figure 9.

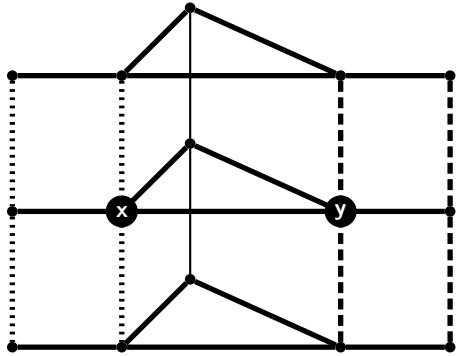

FIGURE 9. Cartesian skeleton of the strong product graph, which has factors induced by one thick and one dashed colored component. Application of Algorithm 1 identifies Cartesian edges in three distinct color classes indicated by thick lines and the two types of dashed lines. The edges drawn as thin lines are not identified as Cartesian because they do not satisfy the *S1-condition*. The backbone of $G$ consists of the vertices $x$ and $y$.

It remains to show how we can identify colors of different colored $G_i$-fibers. For this the next lemma is crucial.

**Lemma 5.3 (Square Lemma [6]).** *Let $G$ be a Cartesian product. If $e$ and $f$ are incident edges of different fibers, then there exists exactly one square without diagonals that contains $e$ and $f$.*

*Furthermore any two opposite edges of a diagonal-free square are edges from copies of one and the same factor.*

In the following, we investigate how we can find the necessary squares and under which conditions we can identify colors of differently colored fibers that belong to one and the same factor.

Before stating the next lemma, we explain its practical relevance. Let $G = \boxtimes_{l=1}^{n} G_l \in \Upsilon$ be a strong product graph. In this case, different fibers for the same factor may be colored differently, see Figure 9 for an example. We will show that in this case there is a square of Cartesian edges containing one Cartesian edge of each of the fibers $G_i^a$ and $G_i^x$, if these fibers are connected by an arbitrary Cartesian edge of some $G_j$-fiber. The other two Cartesian edges then belong to two distinct $G_j$-fibers $G_j^A$ and $G_j^B$, both of which satisfy the *S1-condition*. The existence of such a square implies that $G_i^a$ and $G_i^x$ are copies of the same factor. Thus we can identify the fibers that belong to the same factor after computing a proper horizontal fiber coloring as explained in previous subsection. Moreover we will show in Lemma 5.5 that all parallel fibers that satisfy the *S1-condition* are connected by path of Cartesian edges. This provides that we can color *all* $G_i$-fibers with the same color applying Lemma 5.3 and 5.4.

**Lemma 5.4.** *Let $G = \boxtimes_{l=1}^{n} G_l$ be the strong product of thin graphs. Let there be two different fiber $G_i^a$ and $G_i^x$ that satisfy the* S1-condition.
*Furthermore let there exist an index $j \in \{1, \dots n\}$ s.t. $(p_j(a), p_j(x)) \in E(G_j)$ and $p_k(a) = p_k(x)$ for all $k \neq i, j$.*
*Then there is a square $A\widehat{A}\widehat{B}B$ in $G$ with*

1. $(A, \widehat{A}) \in E(G_i^x)$ *and* $(B, \widehat{B}) \in E(G_i^a)$ *and*
2. $(A, B) \in E(G_j^A)$ *and* $(\widehat{A}, \widehat{B}) \in E(G_j^{\widehat{A}})$, *whereby $G_j^A \neq G_j^{\widehat{A}}$ and at least one edge of $G_j^A$ and at least one edge of $G_j^{\widehat{A}}$ satisfies the* S1-condition.

*Proof.* Since the strong product is commutative and associative it suffices to show this for the product $G = G_1 \boxtimes G_2 \boxtimes G_3$ of thin (not necessarily prime) graphs. W.l.o.g. choose $i = 1$, $j = 2$ and $k = 3$. W.l.o.g., let $x$ have coordinates $(x_1, x_2, x_3)$ and $a$ have coordinates $(a_1, a_2, x_3)$. Now we have to distinguish the following cases for the three different graphs.

Before we proceed we fix a particular notation for the coordinates of certain vertices and edges, which we will maintain throughout the rest of the proof.

- For $G_1$:
    1. $|\mathbb{B}(G_1)| > 1$, i.e. there is an edge $(v_1, \hat{v}_1) \in E(G_1)$ with $v_1, \hat{v}_1 \in \mathbb{B}(G)$ and
    2. not (1): $|\mathbb{B}(G_1)| = |\{v_1\}| = 1$.
- For $G_2$:
    A. the edge $(a_2, x_2)$ satisfies the *S1-condition* in $G_2$
    B. not (A).
    Notice that in case (A) there is by definition a vertex $z_2 \in N[a_2] \cap N[x_2]$ with $|S_{z_2}(a_2)| = 1$ or $|S_{z_2}(x_2)| = 1$. In the following we will assume w.l.o.g. that in this case $|S_{z_2}(x_2)| = 1$.
    Case (B) implies that $|S_{x_2}(x_2)| > 1$. By Theorem 2.13 we can conclude that there is a vertex $\tilde{x}_2 \in N[x_2]$ with $|S_{\tilde{x}_2}(\tilde{x}_2)| = 1$, which implies that the edge $(x_2, \tilde{x}_2)$ satisfies the *S1-condition* in $G_2$.
- For $G_3$:
    i. $x_3 \in \mathbb{B}(G_3)$
    ii. not (i): $x_3 \notin \mathbb{B}(G_3)$.
    For the sake of convenience define $\tilde{p}_3 = x_3$ if we have case (i). In case (ii) let $\tilde{p}_3 = z_3$ with $z_3 \in N[x_3]$ s.t. $|S_{z_3}(x_3)| = 1$. Notice that such a vertex $z_3$ has to exist in $G_3$, since otherwise $|S_{z_3}(x_3)| > 1$ for all $z_3 \in N[x_3]$. But then for all $z, x \in V(G)$ with $z \in N[x]$ with coordinates $z = (\ , \ , z_3)$ and $x = (\ , \ , x_3)$, resp., holds $|S_z(x)| = \prod_{i=1}^{3} |S_{z_i}(x_i)| > 1$. Hence none of the edges of $G_1^a$ and $G_1^x$ satisfies the *S1-condition*, contradicting the assumption. However, notice that $\tilde{p}_3$ is chosen such that $|S_{\tilde{p}_3}(x_3)| = 1$.

In all cases we will choose the coordinates of the vertices of the square $A\widehat{A}\widehat{B}B$ as follows: $A = (v_1, x_2, x_3)$, $B = (v_1, a_2, x_3)$ with $v_1 \in \mathbb{B}(G_1)$ and $\widehat{A} = (\hat{v}_1, x_2, x_3)$, $\widehat{B} = (\hat{v}_1, a_2, x_3)$, $v_1 \neq \hat{v}_1$. By choice holds $(A, \widehat{A}) \in G_1^x$, $(B, \widehat{B}) \in E(G_1^q)$, $(A, B) \in G_2^A$ and $(\widehat{A}, \widehat{B}) \in E(G_2^{\widehat{A}})$ whereby $G_2^A \neq G_2^{\widehat{A}}$, see Figure 10.

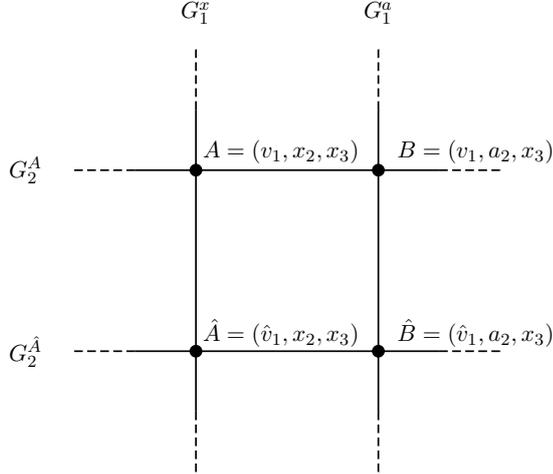

FIGURE 10. General notation of the chosen square $A\widehat{A}\widehat{B}B$.

It remains to show that at least one edge of both fibers $G_2^A$ and $G_2^{\widehat{A}}$ satisfies the *S1-condition*. This part of the proof will become very technical. In Figure 11 and 12 the ideas of the proofs are depicted.

**Cases 1.A.i and 1.A.ii :**
Let $v_1, \hat{v}_1 \in \mathbb{B}(G_1)$ with $(v_1, \hat{v}_1) \in E(G_1)$. Let $z_2 \in N[x_2]$ with $|S_{z_2}(x_2)| = 1$ in $G_2$. Choose $z \in V(G)$ with coordinates $(v_1, z_2, \tilde{p}_3)$.

By definition of the strong product the edges $(z, A)$ and $(z, B)$ do exist in $G$ and therefore $z \in N[A] \cap N[B]$. Moreover Corollary 2.8 implies that $|S_z(A)| = 1$. Therefore the edge $(A, B)$ is satisfying the *S1-condition* in $G$ in both cases (i) and (ii).

The same argument holds for the edge $(\widehat{A}, \widehat{B})$ by choosing $z \in V(G)$ with coordinates $(\hat{v}_1, z_2, \tilde{p}_3)$.

**Case 1.B.i and 1.B.ii :**
Let $v_1, \hat{v}_1 \in \mathbb{B}(G_1)$ with $(v_1, \hat{v}_1) \in E(G_1)$ and let $\tilde{x}_2 \in N[x_2]$ with $|S_{\tilde{x}_2}(\tilde{x}_2)| = 1$. Choose $z \in V(G)$ with coordinates $(v_1, \tilde{x}_2, \tilde{p}_3)$. By definition of the strong product holds that $(z, A) \in E(G)$.

In case (i) we can conclude from Corollary 2.8 that $|S_z(z)| = 1$. Moreover in case (i) holds by definition of the strong product that $(z, A) \in G_2^A$ and we are ready.

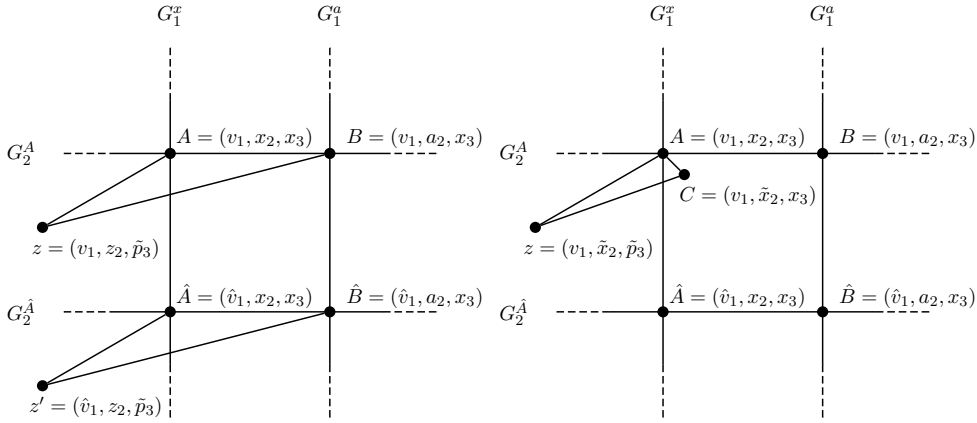

Figure 11. Left: Case 1.$A.i.$ and $ii.$. Right: Case 1.$B.i.$ and $ii.$

Otherwise in case (ii) choose the vertex $C$ with coordinates $(v_1, \tilde{x}_2, x_3)$. Then $z \in N[A] \cap N[C]$ and $|S_z(C)| = 1$. Since $(A, C) \in G_2^A$ the assumption for $G_2^A$ follows.

The same arguments hold for $G_2^{\hat{A}}$ by choosing $z \in V(G)$ with coordinates $(\hat{v}_1, \tilde{x}_2, \tilde{p}_3)$ and $C$ with coordinates $(\hat{v}_1, \tilde{x}_2, x_3)$.

**Cases 2.A.i and 2.A.ii :**

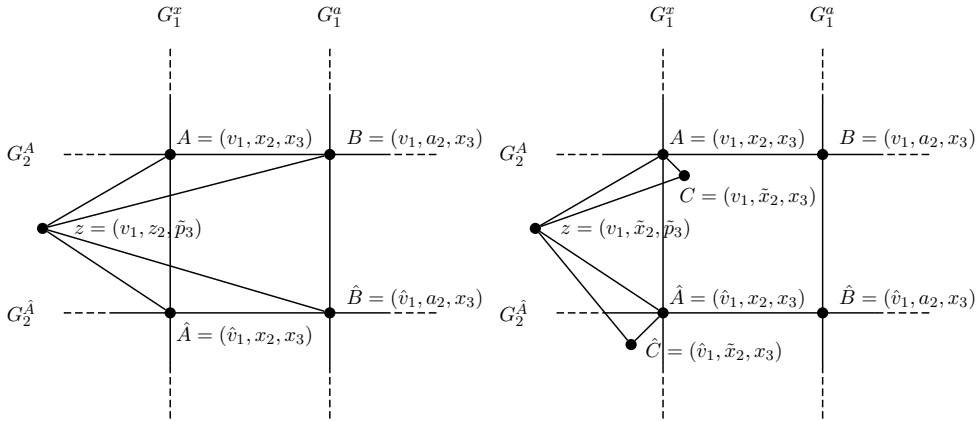

Figure 12. Left: Case 2.$A.i.$ and $ii.$. Right: Case 2.$B.i.$ and $ii.$

Let $v_1 \in \mathbb{B}(G_1)$ and $\hat{v}_1 \in N[v_1]$. Let $z_2 \in N[x_2] \cap N[a_2]$ with $|S_{z_2}(x_2)| = 1$ in $G_2$. Choose $z \in V(G)$ with coordinates $(v_1, z_2, \tilde{p}_3)$.

In order to show that the conditions are fulfilled for $G_2^A$ we proceed as in cases in (1.A.i) and (1.A.ii):

By definition of the strong product there are non-Cartesian edges $(z, \widehat{A})$ and $(z, \widehat{B})$ and thus $z \in N[\widehat{A}] \cap N[\widehat{B}]$. Now, Lemma 3.6 implies that $|S_{v_1}(\hat{v}_1)| = 1$ and therefore by applying Corollary 2.8 we can conclude that $|S_z(\widehat{A})| = 1$, and the assumption follows for $G_2^{\widehat{A}}$.

**Case 2.B.i and 2.B.ii :**
Let $v_1 \in \mathbb{B}(G_1)$ and $\hat{v}_1 \in N[v_1]$. Let $\tilde{x}_2 \in N[x_2]$ with $|S_{\tilde{x}_2}(\tilde{x}_2)| = 1$ Choose $z \in V(G)$ with coordinates $(v_1, \tilde{x}_2, \tilde{p}_3)$.

That the conditions are fulfilled for $G_2^A$ is shown analogously, as in cases in (1.B.i) and (1.B.ii).

To show that the conditions are also fulfilled in case (2.B.i) and (2.B.ii) for $G_2^{\widehat{A}}$ choose $z \in V(G)$ with coordinates $(v_1, \tilde{x}_2, \tilde{p}_3)$ and a vertex $C$ with coordinates $(\hat{v}_1, \tilde{x}_2, x_3)$. Clearly $(\widehat{A}, \widehat{C}) \in E(G_2^{\widehat{A}})$. Furthermore, by definition of the strong product, $z \in N[\widehat{A}]$ and $z \in N[\widehat{C}]$, and thus $z \in N[\widehat{A}] \cap N[\widehat{C}]$. By applying Corollary 2.8 we now conclude that $|S_z(\widehat{C})| = 1$, using that Lemma 3.6 implies $|S_{v_1}(\hat{v}_1)| = 1$. Thus the edge $(\widehat{A}, \widehat{C})$ satisfies the *S1-condition*, and the assumption follows for $G_2^{\widehat{A}}$. □

It is important to notice that the square $A\widehat{A}\widehat{B}B$ in the construction of Lemma 5.3 is composed exclusively of Cartesian edges. The lemma can therefore be applied to determine whether two fibers $G_i^a$ and $G_i^x$, which have been colored differently in the initial steps, are copies of the same factor, and hence, whether their colors need to be identified. As we shall see below, this approach is in fact sufficient to identify all fibers belonging to a common factor.

**Lemma 5.5.** *Let $G = \boxtimes_{j=1}^n G_j$ be the strong product of thin graphs. Furthermore let $G_i^{y_1}, \ldots, G_i^{y_m}$ be all $G_i$-fibers in $G$ satisfying the* S1-condition. *Then there is a connected path $\mathcal{P}$ in $G$ consisting only of vertices of $\mathcal{X} = \{x_1, \ldots, x_m\}$ with $x_j \in V(G_i^{y_j})$ s.t. each edge $(x_k, x_l) \in \mathcal{P}$ is Cartesian.*

*Proof.* Since the strong product is commutative and associative it suffices to show this for the product $G = G_1 \boxtimes G_2$ of two thin (not necessarily prime) graphs. W.l.o.g. let $i = 1$. Moreover, we can choose w.l.o.g. the vertices $x_1, \ldots, x_m$ such that $p_1(x_k) = x$ for $k = 1, \ldots, m$. Moreover by applying Theorem 2.13 we can choose $x$ such that $x \in \mathbb{B}(G_1)$.

Consider first all vertices $v$ with coordinates $(x, v_2)$ such that $v_2 \in \mathbb{B}(G_2)$. From Theorem 2.13 follows that $\mathbb{B}(G_2)$ is connected. Thus there is connected path $\mathcal{P}_2$ consisting only of such vertices $v$. Moreover each edge $(a, b)$ with $a, b \in V(\mathcal{P}_2)$ and thus with coordinates $(x, a_2)$ and $(x, b_2)$, resp., is Cartesian. Furthermore all corresponding $G_1^v$-fibers are satisfying the *S1-condition*, since for each edge $(v, w)$ holds $|S_v(v)| = 1$ , by applying Corollary 2.8. Therefore all vertices $v$ with coordinates $(x, v_2)$ with $v_2 \in \mathbb{B}(G_2)$ are also contained in $\mathcal{X}$. Hence all those $G_i^v$-fibers are connected by such a path $\mathcal{P}_2$ with $V(\mathcal{P}_2) \subset \mathcal{X}$.

Let now $\tilde{v}$ be any vertex in $\mathcal{X} \backslash V(\mathcal{P}_2)$. Hence $p_2(\tilde{v}) \notin B(G_2)$. Theorem 2.13 implies that for all those vertices $p_2(\tilde{v}) \notin B(G_2)$ there is an adjacent vertex $p_2(v)$ in $G_2$ s.t. $p_2(v) \in \mathbb{B}(G_2)$. Thus we can conclude that for all vertices $\tilde{v} \in \mathcal{X} \backslash V(\mathcal{P}_2)$ with coordinates $(x, p_2(\tilde{v}))$ there is an adjacent vertex $v \in V(\mathcal{P}_2)$ with coordinates $(x, p_2(v))$, from what the assumption follows. □

### 5.3. Detection of unidentified Cartesian edges

One question still remains open: How can we identify a Cartesian $(x, y)$ edge that does not satisfy the *S1-condition*, i.e., if for all $z \in N[x] \cap N[y]$, we have both $|S_z(x)| > 1$ and $|S_z(y)| > 1$? Figures 13 and 14 show examples of product graphs, in which not all fibers were determined by the approach outline in the previous two sections.

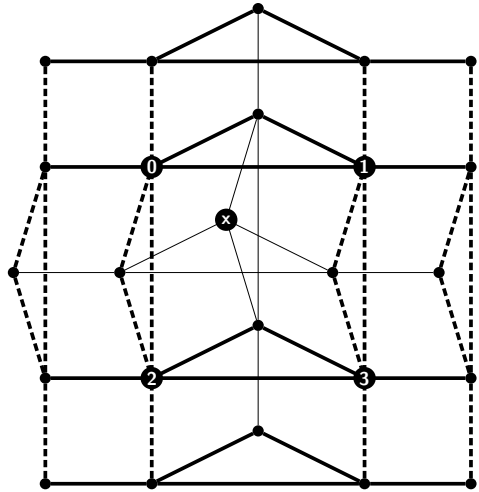

FIGURE 13. Cartesian Skeleton of the strong product $G$ of two prime factors induced by the dashed and bold lined fibers. Application of Algorithm 1 to all fibers determines a part of the Cartesian Skeleton $H$ that consists only of the edges drawn as dashed or bold lines. While the bold and dashed fibers identify the true factors, we miss the copies shown by thin lines. None of these edges satisfies the *S1-condition* in an induced 1-neighborhood. The backbone $\mathbb{B}(G)$ consists of the vertices 0, 1, 2 and 3.

Unfortunately, we do not see an efficient possibility to resolve the missing cases by utilizing only the information contained in the fibers that already have been identified so far and the structure of 1-neighborhoods. We therefore resort to a "brute-force" method which relies on the identification of Cartesian edges within 2-neighborhoods.

Of course, it would be desirable if smaller structure were sufficient. Natural candidates would be to exploit the *S1-condition* in unions of adjacent neighborhoods of the form $\langle N[x] \cup N[x'] \rangle$, where $(x, x')$ is a Cartesian edge. Such subgraphs are also sub-products, see [5]. However, the example in Figure 14 shows that the information contained in these subproducts is still insufficient.

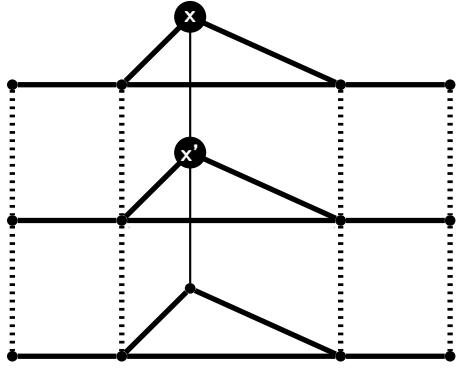

FIGURE 14. Cartesian skeleton of a thin strong product graph whose factors are induced by one thick and dashed component. The fiber whose edges are drawn as thin lines does not satisfy the *S1-condition*. Moreover, even in the subgraph induced by the neighborhoods of $x$ and $x'$, which is the product of a path and a $K_3$, the *S1-condition* is violated for the Cartesian edge.

In the following we will show that every Cartesian $(x, y)$ edge that does not satisfy the *S1-condition* can be determined as Cartesian in the 2-neighborhood $N_2[x]$.

**Lemma 5.6.** *Let $G = \boxtimes_{j=1}^n G_j$ be the strong product of thin graphs. Let $(x, x')$ be any Cartesian edge of $G$. Then $|S_{\langle N_2[x] \rangle}(x)| = 1$, i.e., the edge $(x, x')$ satisfies the* S1-condition *in $\langle N_2[x] \rangle$.*

*Proof.* Assume $|S_{\langle N_2[x] \rangle}(x)| > 1$. Then there is a vertex $v \neq x$ with $N[v] \cap N_2[x] = N[x] \cap N_2[x]$. This would mean that $N[v] = N[x]$ in $G$, since $N[v] \cap N_2[x] = N[v]$ and $N[x] \cap N_2[x] = N[x]$, contradicting that $G$ is thin.   □

We will show in the next lemma that the PFD of an arbitrary 2-neighborhood $N_2$ is not finer than the PFD of a given graph $G \in \Upsilon_n$. This implies that each Cartesian edge in $G$ that is contained in $N_2$ and satisfies the *S1-condition* in $N_2$ can be determined as Cartesian in $N_2$.

**Lemma 5.7.** *Let $G \in \Upsilon_n$ and let $x$ be an arbitrary vertex in $V(G)$. Then $|PF(\langle N_2[x] \rangle)| = n$.*

*Proof.* Notice that $|PF(G)| = n$ and $|PF(N[x])| = n$, since $G \in \Upsilon_n$. Lemma 2.2 implies that the PFD of any neighborhood (2 - neighborhoods included) in a graph $G$ has at least $|PF(G)|$ factors. Applying this fact two times we have

$$n = |PF(G)| \leq |PF(N_2[x])| \leq |PF(N[x])| = n,$$

and thus $|PF(H)| = n$. □

From Lemma 5.6 and 5.7 we can conclude that any Cartesian edge $(x, y)$ of some fiber that does not satisfy the *S1-condition* can be determined as Cartesian in the 2-neighborhood $\langle N_2[x] \rangle$. Thus we can identify *all* Cartesian edges of $G$.

The last step we have to consider is to identify such fibers as copy of the corresponding factor. This can be done in a simple way. Consider that we have now identified *all* Cartesian edges of $G$. Notice that for all new identified $G_i^a$-fibers holds $a \notin \mathbb{B}(G)$, otherwise each edge containing vertex $a$ of this fiber would satisfy the *S1-condition* in $\langle N[a] \rangle$ and we would have identified this fiber. But for each such vertex $a$ there is a vertex $x \in N[a]$ with $x \in \mathbb{B}(G)$ since $\mathbb{B}(G)$ is a connected dominating set. Thus the corresponding $G_i^x$-fiber satisfies the *S1-condition* and is therefore already identified and colored as $G_i$-fiber. Hence again we can apply the square property to determine such a new identified $G_i^a$-fiber belonging to a copy of the factor $G_i$ by identifying the colors of the $G_i^a$-fiber with the color of the $G_i^x$-fiber.

### 5.4. Algorithm and Time Complexity

We will now summarize the algorithm for determining the colored Cartesian skeleton of a given graph $G \in \Upsilon$ w.r.t. to its PFD and give the top level control structure, which are proved to be correct in the preceding subsections. Moreover we will determine the time complexity, which is stated in the following lemma.

---

**Algorithm 2** Cartesian skeleton and Product Coloring of $G$

---

1: **INPUT:** Graph $G \in \Upsilon$.
2: Compute the backbone $\mathbb{B}(G)$;
3: **for** all $x$ in $\mathbb{B}(G)$ **do**
4:    Color all $G_i^x$-fibers (and $G_i^y$-fibers that satisfy the *S1-condition*) with Algorithm 1;
5: **end for**
6: Determine unidentified Cartesian edges in $N_2$ neighborhoods;
7: Compute all squares in the induced Cartesian skeleton of $G$ and identify the colors of parallel fibers applying Lemma 5.3;
8: **OUTPUT:** Product coloring of $G$ with respect to its PFD;

---

**Lemma 5.8.** *Algorithm 2 determines the colored Cartesian skeleton with respect to its PFD of a given graph $G = (V, E) \in \Upsilon$ with bounded maximum degree $\Delta$ in $O(|V|^2 \cdot \Delta^{10})$ time.*

*Proof.* **1. Determining the backbone** $\mathbb{B}(G)$**:** we have to check for a particular vertex $v \in V(G)$ whether there is a vertex $w \in N[v]$ with $N[w] \cap N[v] = N[v]$. This can be done in $O(\Delta^2)$ for a particular vertex $w$ in $N[v]$. Since this must be done for all vertices in $N[v]$ we end in time-complexity $O(\Delta^3)$. This step must be repeated for all $|V|$ vertices of $G$. Hence the time complexity for determining $\mathbb{B}(G)$ is $O(|V| \cdot \Delta^3)$.

**2. For-Loop.** The time complexity of Algorithm 1 is $O(|V| \cdot \log_2(\Delta) \cdot (\Delta)^6)$. The for-loop is repeated for all backbone vertices. Hence we can conclude that the time complexity of the for-loop is $O(|V| \cdot |V| \cdot \log_2(\Delta) \cdot \Delta^6)$.

**3. Determine unidentified Cartesian edges in $N_2$ neighborhoods.** Notice that each $N_2$ neighborhood has at most $1 + \Delta \cdot (\Delta - 1)$ vertices. The decomposition of each $N_2$ with the algorithm of Feigenbaum and Schäffer [3] and hence the assignment to an edge of being Cartesian is bounded by $O((1 + \Delta(\Delta - 1))^5)$. Again this will be repeated for all vertices and thus the time complexity is $O(|V| \cdot (1 + \Delta(\Delta - 1))^5) = O(|V| \cdot \Delta^{10})$.

**4. Compute all squares.** Take an edge $(x, y)$ and check whether there is an edge $(x_i, y_j)$ for all neighbors $x_1, \ldots, x_l \neq y$ of $x$ and $y_1, \ldots, y_k \neq x$ of $y$. Notice that $l, k \leq \Delta - 1$. This leads to all squares containing the edge $(x, y)$ and requires at most $(\Delta - 1)^2$ comparisons. Since we need diagonal-free squares we also have to check that there is no (Cartesian) edge $(x, y_j)$ and no edge $(x_i, y)$. This will be done for all $|E|$ edges. Thus we end in time complexity $O(|E| \cdot (\Delta - 1)^3)$, which is $O(|V| \cdot \Delta^4)$, since the number of edges in $G$ is bounded by $|V| \cdot \Delta$.

Considering all steps we end in an overall time complexity $O(|V|^2 \cdot \Delta^{10})$. $\quad\square$

## 6. Recognition of graphs $G \in \Upsilon$

In this section we will provide an algorithm that tests whether a given graph is element of $\Upsilon$ in polynomial time.

**Lemma 6.1.** *Algorithm 3 recognizes if a given graph $G$ is in class $\Upsilon$.*

*Proof.* Lemma 2.2 implies that the PFD of any neighborhood in a graph $G$ has at least $|PF(G)|$ factors and hence $MAX \geq |PF(G)|$. Thus if $MAX = |PF(G)|$ then none of the decomposed neighborhoods was locally finer. If in addition the isomorphism test is true we can conclude that we have found the correct factors and that $G \in \Upsilon$. $\quad\square$

**Lemma 6.2.** *Algorithm 3 recognizes if a given a given graph $G = (V, E)$ with bounded maximum degree $\Delta$ is in class $\Upsilon$ in $O(|V|^2 \cdot \Delta^{10})$ time.*

*Proof.* Algorithmus 2 takes $O(|V|^2 \cdot \Delta^{10})$ time. Computing the maximum $MAX$ of the number of prime factors of each decomposed neighborhood, extracting the possible factors and the isomorphism test for a fixed bijection can be done in linear time in the number of vertices. Thus we end in $O(|V|^2 \cdot \Delta^{10})$ time. $\quad\square$

---

**Algorithm 3** Recognition if $G \in \Upsilon$

---

1: **INPUT:** thin Graph $G$.
2: compute the colored Cartesian skeleton of $G$ with Algorithm 2 and remind the number of prime factors in each decomposed neighborhood;
3: $MAX \leftarrow$ maximal number of prime factors of decomposed neighborhoods;
4: compute the possible prime factors $G_1, \ldots, G_m$ of $G$ by taking one connected component of the Cartesian skeleton of each color $1, \ldots, m$;
5: **if** $\boxtimes_{i=1}^m G_i \simeq G$ and $MAX = m$ **then**
6:    IS_IN_$\Upsilon \leftarrow$ true;
7: **else**
8:    IS_IN_$\Upsilon \leftarrow$ false;
9: **end if**
10: **OUTPUT:** IS_IN_$\Upsilon$;

---

## References

[1] W. Dörfler and W. Imrich. Über das starke Produkt von endlichen Graphen. *Österreih. Akad. Wiss., Mathem.-Natur. Kl., S.-B .II*, 178:247–262, 1969.

[2] J. Feigenbaum. Product graphs: some algorithmic and combinatorial results. Technical Report STAN-CS-86-1121, Stanford University, Computer Science, 1986. PhD Thesis.

[3] J. Feigenbaum and A. A. Schäffer. Finding the prime factors of strong direct product graphs in polynomial time. *Discrete Math.*, 109:77–102, 1992.

[4] Marc Hellmuth, Wilfried Imrich, Werner Klöckl, and Peter F. Stadler. Approximate graph products. *European Journal of Combinatorics*, 30:1119 – 1133, 2009.

[5] W. Imrich and S Klavžar. *Product graphs*. Wiley-Interscience Series in Discrete Mathematics and Optimization. Wiley-Interscience, New York, 2000.

[6] W. Imrich and I. Peterin. Recognizing cartesian products in linear time. *Discrete Math.*, 307:472–482, 2007.

[7] Ali Kaveh and K. Koohestani. Graph products for configuration processing of space structures. *Comput. Struct.*, 86:1219–1231, 2008.

[8] Günter Wagner and Peter F. Stadler. Quasi-independence, homology and the unity of type: A topological theory of characters. *J. Theor. Biol.*, 220:505–527, 2003.

[9] Blaž Zmazek and Janez Žerovnik. Weak reconstruction of strong product graphs. *Discrete Math.*, 307:641–649, 2007.

## Acknowledgements

This work was supported by a collaborative grant of the Austrian *FWF* (Proj No. P18969-N15) and the German *DFG* (Proj No. STA850/5-1). We thank the anonymous reviewer for their constructive comments.

Marc Hellmuth
Bioinformatics Group,
Department of Computer Science; and Interdisciplinary Center for Bioinformatics,
University of Leipzig,
Härtelstrasse 16-18, D-04107 Leipzig, Germany

Max Planck Institute for Mathematics in the Sciences
Inselstrasse 22, D-04103 Leipzig, Germany
e-mail: `marc@bioinf.uni-leipzig.de`

Wilfried Imrich
Chair of Applied Mathematics
Montanuniversität, A-8700 Leoben,
Austria
e-mail: `imrich@unileoben.ac.at`

Werner Klöckl
Chair of Applied Mathematics
Montanuniversität, A-8700 Leoben,
Austria
e-mail: `werner.kloeckl@mu-leoben.at`

Peter F. Stadler
Bioinformatics Group,
Department of Computer Science; and Interdisciplinary Center for Bioinformatics, University of Leipzig,
Härtelstrasse 16-18, D-04107 Leipzig,Germany

Max Planck Institute for Mathematics in the Sciences
Inselstrasse 22, D-04103 Leipzig, Germany

RNomics Group, Fraunhofer Institut für Zelltherapie und Immunologie, Deutscher Platz 5e, D-04103 Leipzig, Germany

Department of Theoretical Chemistry, University of Vienna, Währingerstraße 17, A-1090 Wien, Austria

Santa Fe Institute, 1399 Hyde Park Rd., Santa Fe, NM87501, USA
e-mail: `studla@bioinf.uni-leipzig.de`